# BEYOND STEM, HOW CAN WOMEN ENGAGE BIG DATA, ANALYTICS, ROBOTICS AND ARTIFICIAL INTELLIGENCE? – *AN EXPLORATORY ANALYSIS OF CONFIDENCE AND EDUCATIONAL FACTORS IN THE EMERGING TECHNOLOGY WAVES INFLUENCING THE ROLE OF, AND IMPACT UPON, WOMEN.*

**Abstract:**

In spite of the rapidly advancing global technological environment, the professional participation of women in technology, big data, analytics, artificial intelligence and information systems related domains remains proportionately low. Furthermore, it is of no less concern that the number of women in leadership in these domains are in even lower proportions. In spite of numerous initiatives to improve the participation of women in technological domains, there is an increasing need to gain additional insights into this phenomenon especially since it occurs in nations and geographies which have seen a sharp rise in overall female education, without such increase translating into a corresponding spurt in information systems and technological roles for women. The present paper presents findings from an exploratory analysis and outlines a framework to gain insights into educational factors in the emerging technology waves influencing the role of, and impact upon, women. We specifically identify 'ways for learning' and 'self-efficacy' as key factors, which together lead us to the 'Advancement of Women in Technology' (AWT)  insights framework.  Based on the AWT framework, we also proposition principles that can be used to encourage higher professional engagement of women in emerging and advanced technologies.

**Key Words:** Women's Education, Technology, Artificial Intelligence, Knowing, Confidence, Self-Efficacy, Learning.

**Introduction**

*"I think it's very important to get more women into computing. My slogan is: Computing is too important to be left to men."*

~ Karen Spärck Jones, Professor, Cambridge Computer Laboratory.

Women have been significantly underrepresented in scientific, technological and quantitative domains over the past few decades and in their reasonably comprehensive study "Why so few? Women in science, technology, engineering, and mathematics" (Hill, Corbett, Rose, 2010), the authors advocate a proactive approach to cultivating an early interest in science, technology, engineering, and mathematics (STEM) disciplines and in articulating an inclusive STEM supportive environment for women in their educational settings. There are a variety of serious concerns that have been raised surrounding the low proportion of women in STEM. This proportion bias phenomenon reflects an underutilization of human capital, which has socioeconomic consequences, along with implications for educational frameworks for academia as well for employment frameworks in industry (Ong, Wright, Espinosa, & Orfield, 2011).

Global leaders, corporations, educational institutions and the world at large acknowledge the tremendous benefits of educating girls and women, especially in STEM disciplines and in providing a supportive environment for women in STEM associated professions. Yet, the reality is that the presence of women in education, research and in practice in computer science and STEM domains remains low. Additionally this plight is further compounded by the fact that even fewer women reach leadership positions in these domains, a reality aptly captured in the adage 'the higher up you look, the fewer women you see'. According to the National Girls Collaborative Project, women make up half of the total college educated workforce, however "only 29% of the science and engineering workforce." Some statistics that they provide are as follows: 35% of chemists are women, 11% of physicists and astronomers, 22.7% chemical engineers, 17.5% architectural engineers, 10% are computer hardware engineers. "Minority women comprise fewer than 1 in 10 employed scientists and engineers" (ngcproject.org). Although the percentage of women in male dominated fields has increased, the disparity is still staggering. Interestingly, recent trends show that ""Female and male students enrolled in advanced science courses at comparable rates, with females slightly more likely than males to do so (22% versus 18%)" (http://ngcproject.org/statistics) – however the same 2016 report also shows that though math and calculus enrollments did not show significant gender based differences, yet "Male students were more likely than female students to take engineering (3% versus 1%) and computer science courses (7% versus 4%) and enrolled in AP computer science A at a much higher rate (81% males; 19% females)".

Much research has already been done on the topic of underrepresentation of women in STEM disciplines highlighting a variety of associated issues and factors (Diekman, Clark, Brown & Johnston, 2017; Bonham & Stefan, 2017; Tully, 2017; Daldrup-Link, 2017). Numerous useful recommendations have been provided and initiatives and programs have been implemented (Bystydzienski & Bird, 2006; Young, Young & Paufler, 2017; Katz, et. al., 2017; Hill & Rose, 2010). To the best our knowledge, our research is unique as we look into the future beyond STEM and into the emerging technological ecosystem immersed in Big Data, Analytics, Machine and Deep Learning, Artificial Intelligence and IoT, and explore the role of and impact upon women, with a lead research inquiry:

"How can women engage the big data, analytics, machine and deep learning, robotics and artificial intelligence wave?"

It is absolutely necessary to grasp the core difference between a generic STEM education, which is of itself of great importance, and the even more critical dimension of education and immersion into emerging technologies – this difference is elaborated upon in the literature review and subsequent sections of this study. Our research focuses on conducting an exploratory analysis of educational factors influencing the role of, and impact upon, women in the context of Big data, Machine learning, AI & related futuristic technologies.

**Literature Review:**

In the pursuit of progress and scientific advancement, humanity has developed numerous inventions and theories – a centerpiece of which in recent decades has been waves of technological innovations. Past developments of technology have created efficiencies which reduced the need for human muscle power. The information age facilitated humankind's dependence on commuting technologies for storing, processing and communicating information. Thus the technologies that we developed till recently allowed humankind to still remain 'on top', ruling over machines that did our bidding. Past and present STEM education for women and technological jobs for women movements were designed to address such technologies, where humankind intelligently took all decisions and controlled machines to do their bidding (Hill, C., Corbett, C., & St Rose, A., 2010; Katz, L. A., et al., 2017).

It is critical to note the characteristics of the present technological wave that has begun to emerge: 'Big Data, Analytics, Robotics and Artificial Intelligence'. 'Big Data' refers to the unprecedented nature of current data - vast quantities of data being generated, with variety (types of data), velocity (the speed at which data is being generated or transformed) and veracity (uncertainty) in high measure. This data is being used by industries and governments to gain insights, take data driven decisions and create value. This also creates challenges for privacy and individual rights, businesses, nature of society and its governance (Chen, H., Chiang, R. H., & Storey, V. C., 2012; O'Neil, C., 2017; Drosou, M., Jagadish, H. V., Pitoura, E., & Stoyanovich, J., 2017; LaValle, S., Lesser, E., Shockley, R., Hopkins, M. S., & Kruschwitz, N., 2011).

The science of robotics opens up possibilities for sophisticated automation and large scale replacement of human beings in a variety of activities. It also creates options for for substantial augmentation of human capabilities (Brynjolfsson, E., & Mcafee,A., 2017; Yang, G. Z., et al., 2016; Mataric, M., 2017). Artificial Intelligence (AI) technologies have vast disruption potential and possess power to outperform human beings on multiple intellectual dimensions – some of the recent developments provide machines with 'learning' capabilities, wherein machines are able to develop programs and insights based its ability to process vast quantities of data, improving iteratively using sophisticated algorithms (Olson, R. et al., 2017; Brynjolfsson, E., & Mcafee, A., 2017; Helbing, D. et al., 2017; Moulin-Frier, C. et al., 2017). The implications are enormous – traditionally, we developed machines and coded machine capabilities on a case-specific basis. In contrast, today we have developed technologies which can self-improve, self-learn, adapt dynamically and thus have increasing degree of autonomy, and proficiencies independent of human control.

This significantly alters the challenge of educating and supporting of women in technological domains. The emphasis of women-in-STEM thus far has been to create solutions which required active human management of technologies. If these solutions are implemented without adaption they will not sufficiently prepare women for the future dominated by big data, robotics, AI and similar technologies which have distinct characteristics, have been fostered in mostly masculine environments and are increasingly autonomous in nature. Supporting the education and professional participation of women in such emerging technologies, that create solutions that replace human beings, requires distinctly new insights and the discussion that follows initiate explorative ideation to address the same.

*Women-Learning*

*"Investing in women is single most effective antidote to the world's pressing problems: war, poverty, disease. Woman plays a special role in society by contributing not only to family wellbeing, but to community wellbeing as a whole."* ~ Global Fund for Women.

"Women as Learners" explores the idea that women learn differently from men. There is a significant body of research exploring gender sensitive learning and women's "ways of knowing" and gender specific differences (Belenky, M. F., et. al., 1986; Brown, L.M., & Gilligan, C.,1992). Past research has also shown that though the ways of learning and knowing differ, yet the performance remains comparable in certain settings (Zhang, Y. Y., Nayga Jr, R. M., & Depositario, D. P. T., 2015), and differs in other settings (Astur, R. S., Ortiz, M. L., & Sutherland, R. J.,1998) highlighting the need to better understand underlying factors and implications for practice. It is also notable that relatively fewer educators and researchers have used the insights associated with idea of women as learners, especially in STEM domains where there has been a focused effort on women's education. We see this gap as an opportunity and a strong need to further research women's ways of learning. Though questions have often surrounded the dynamics of how men and women learn differently, it has been argued that it is highly probable for men and women to learn differently, have different opportunities for learning, and different approaches to learning (Belenky, M. F., et. al., 1986; Barbour, K., 2016). However, having these differences in ways of knowing does not carry an implication that women's ways of knowing are inferior to men's, nor does it mean that women's ways of knowing is superior to men's (Hayes and Flannery, 2000).

Women's learning cannot be understood unless social context in which learning took place is taken into account. These contexts sometimes offer conflicting and complex opportunities for women. We, as educators, need to develop a greater awareness of the social dimension of learning in formal education taking into account other contexts of learning in which women not only learn skills or lessons about themselves, but also how women view themselves as learners and shape their future experiences (Hayes, E., & Flannery, D. 2000).

Sometimes women view themselves as distinct learners and other times they acknowledge the gender neutral common ground as learners. Very often, women are simply unaware of their learning needs and distinct educational wants. Various experiences influence women's identities as learners. As aptly stated: "Nevertheless, such settings influence but do not determine women's identities as learners. Women are actively engaged in reinventing these identities, just as they continually reconstruct other aspects of their identities." (Hayes, E. & Flannery, D. 2000). We, as educators, need to help women become aware of the learning that takes place in and outside of educational institutions, "validate this learning and connect it to the classroom learning experience."(Hayes, E. & Flannery, D. 2000). The way women learn is constantly influenced by their experiences which also influences their self-esteem. Women may not be aware of how gender and culture influence learning as well as how they affect their identities and self-esteem. 'Voice as identity' emphasizes that a key dimension of learning is how women develop and express identities. 'Voice as power' reflects women's desire to acquire their individual and collective power "through expression and validation of their interests, needs, and experiences." . Educators need to choose for themselves which meaning of voice and pattern of talk in group learning situations they want to become more aware of in their teaching and learning. Various meanings of the terms have emerged in professional literature and from women's narratives. Women's connections with themselves include the concepts of global processing, subjective knowing, and intuition (Hayes, E. and Flannery, D.t 2000).

There are numerous applications of the knowledge about women's ways of knowing to adult education practices. One useful area of application is that this knowledge can help educators take more informed actions. There are various explanations for the connected nature of women's learning which include physiological, psychological, sociological, anthropological explanations. One in particular I find very intriguing and that is the physiological explanation. For example, Carl Sagan's research concentrated on the brain and he found that the corpus callosum, which connects the two hemispheres of the brain, is actually larger in women. Today neuroscientists believe that because of the larger connection between the two hemispheres, women use more of the brain at one time when completing a motion or engaging in solving problems. Sagan (1998, p. CI) also states that although there is no hard evidence, " the larger connector may also account for a woman's tendency to exhibit greater intuition." So women are able to follow several trains of thought, whereas men "tend to be focusing intensely on single topics."(Sagan, 1998). The expectation here is that these different meanings of connectedness will add more understanding of women's learning and also enhance further research. We do not claim that men's ways of knowing stand in opposition to women's and past research shows that the differences could also be related to the subject matter addressed (Zhang, Y. Y., Nayga Jr, R. M., & Depositario, D. P. T., 2015; Astur, R. S., Ortiz, M. L., & Sutherland, R. J.,1998). These highlights from past research provide sufficient ground for researchers and adult educators is to "seek out a more complex understanding of this influence." A question that we can ask ourselves is how curriculum can be altered to reflect gender-influenced learning. As we begin to integrate this knowledge and awareness into our teaching practice, many presently male-dominant domains stand to benefit greatly as more women will enter those fields bringing fresh perspectives, insights and knowledge value additions. However, it requires a collaborative action for this transformation to take place, even as it is fairly obvious that it does need to take place(Watts ,2015; Tomlinson, 2014).

| Gender-Category ~ Educational Qualification | Yr: 2013 | Yr: 2015 | % Change |
|---|---|---|---|
| Women Leaders /Practitioners ~ CIS % | 78% | 85% | 8.97% |
| Women Leaders /Practitioners ~ Engineering % | 19% | 28% | 47.37% |
| Women Leaders /Practitioners ~ Business % | 39% | 26% | -33.33% |
| Men Leaders /Practitioners ~ CIS % | 92% | 95% | 3.26% |
| Men Leaders /Practitioners ~ Engineering % | 40% | 43% | 7.50% |
| Men Leaders /Practitioners ~ Business % | 24% | 24% | 0.00% |

Table 1: Change in Educational Gender-Category ~ Educational Qualification

Jane Hugo in *Women as Learners* says (Hayes, Hugo, et. al., 2000) state: "We need to recognize that it's easier to think about women as learners than to do something with our knowledge." It is very important to keep the discourse on the topic of underrepresentation of women in male-dominant fields open and current. More research needs to be conducted and actionable questions need to be articulated and addressed. Helping women secure their place in computer science, information systems, analytics and artificial intelligence, will signify that education is equally accessible and supportive to everyone, irrespective of gender. As women's presence increases in technological domains, the benefits of educating girls and women will be more apparent as through the dynamics of fairness, diversity and balance, they will make this world a better place.

*Confidence & Self –efficacy*

Confidence and self-efficacy have been identified as critical factors in encouraging women towards education and career in STEM disciplines (Hill & Rose, 2010). Confidence has been identified as an important factor in academic success for students (Colbeck., Cabrera & Terenzini, 2001). Past research has also highlighted the lack of confidence as one of main barriers to women engaging and succeeding in technological domains (Howe-Walsh & Turnbull, 2016). However, we need a deeper understanding of the constitution and antecedents of confidence and self-efficacy.

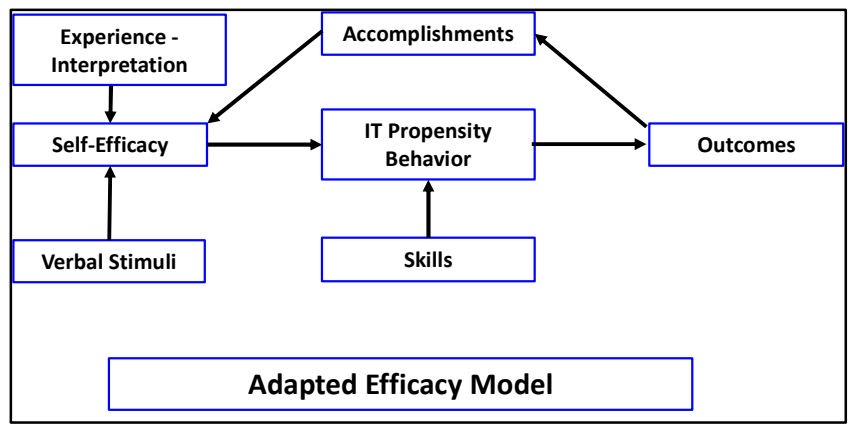

Figure 1: Adapted Efficacy Model (AEM)

Using Bandura's (Bandura, 1977) model of unifying theory of behavioral change, we use inductive logic to proposition an 'adapted efficacy model' to specifically address the women's self-efficacy towards education in and practice of technology disciplines. Self-efficacy is a critical determinant of behavior, and therefore, the interpretation of experience and verbal stimuli become important factors as they influence self-efficacy. Verbal stimuli serves a motivational purpose and is can be exogenous to the person, while interpretation of experiences tends to be endogenous – thus, any attempt to improve the engagement and performance behavior of women must address these two factors (Bandura & Adams, 1977). Education improves skills, which also impacts the behavior and the outcomes, which in creates a sense of accomplishments. Accomplishments have a positive impact on self-efficacy and are distinguished from experiences as being objective measurable peak points attested by external agents, while experiences are viewed from a subjective framework. Though the 'Adaptive Efficacy Model' (AEM) is fundamentally gender neutral in its applicability, the factors in the model are sensitive to gender differences, highlighting the previous research supporting gender differences in self-efficacy towards technology (Scherer & Siddiq, 2015). Past research has highlighted that a "psychological sense of masculinity" influences self-efficacy, not biological gender (Huffman, Whetten & Huffman, 2013). This has important implications for gender sensitivity in the methods used to apply factors that influence self-efficacy.

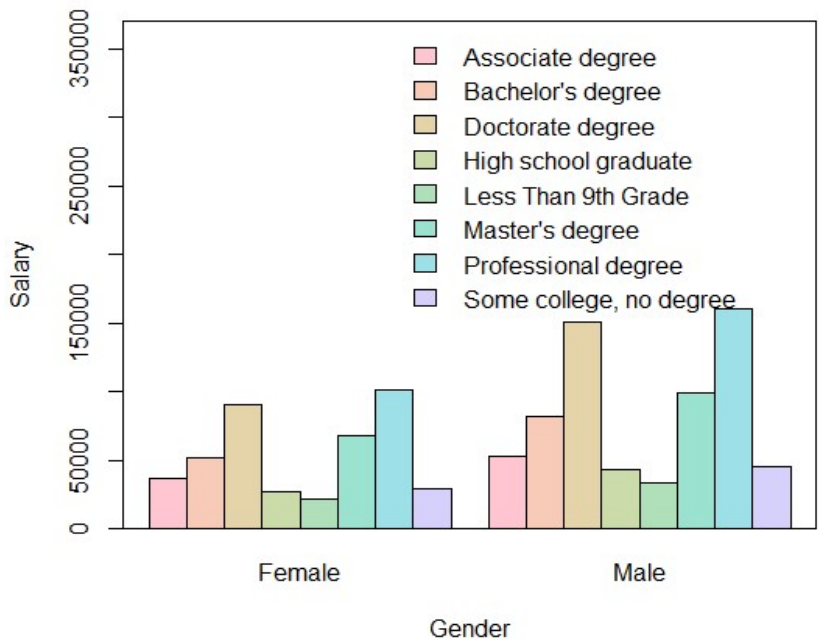

**Figure 2. Mean Salary By Education**

**Methodology, Data & Analysis**

We use inductive reasoning based on literature review and analysis of current technological trends in conjunction with an exploratory analysis on readily available data from the NSF and Statista to identify critical factors and posit the AEM Model (above) and the AWT framework (below). In exploring available data from Statista, we observe (Table 1) a growing interest in technology and engineering education and a drop in business enrollment for women. We also observe that the mean salaries for women continue to remain lower related to their male counterpart for the same levels of education. Due to the paucity of data for direct measurement of the engagement of women in analytics, robotics and artificial intelligence education and practice, we used data from a study on the use of artificial intelligence devices by gender to identify gender differences by confidence and comfort levels.

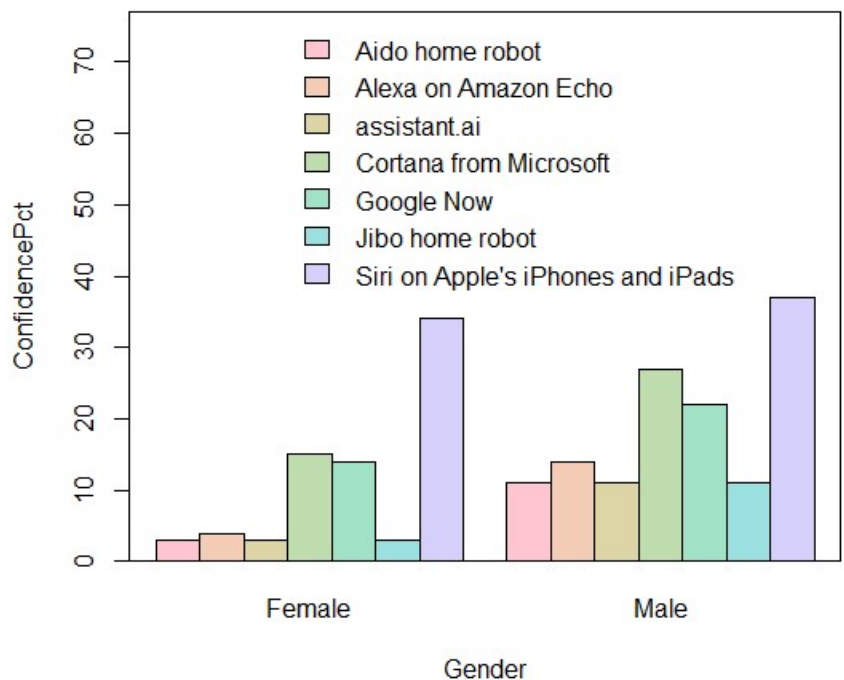

**Figure 2: Confidence in Using AI technologies By Gender**

Based on the analysis of the data obtained from the "Familiarity with virtual digital assistants (VDAs) in the United States, as of September 2016, by gender" survey, we are able to use explorative analysis to indicate that men have greater confidence and self-efficacy with artificial intelligence - based devices (Figure 2.). This highlights a focused need for further study of associated phenomena and underpins the importance of the present research which posits that there is a need to develop self-efficacy using the AEM and AWT frameworks presented in this paper.

**Discussion & Propositions**

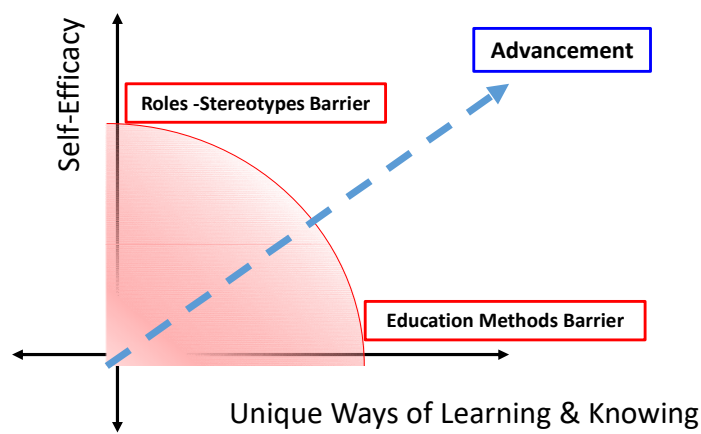

**AWT Framework**
Advancement of Women in Technology Framework

Figure 3: Advancement of Women in Technology (AWT) Framework

It becomes fairly obvious that there is a need to specifically address AI education for women specifically in the context of self-efficacy and gender sensitive learning. We need to address issues that can narrow the education and confidence gap. We need to explore teaching and learning practices that can be implemented to better reflect women's ways of learning. Key questions remain as to how can learning practices, pedagogy, academic frameworks and the educational ecosystem foster a supportive environment for female education? Using inductive logic from literature review, examination of women's ways of learning, exploratory analysis of secondary data and logical progression using fair proxies for AI usage confidence, we develop the following propositions which are reflected in the AWT framework presented below. We juxtapose identified trends in education with trends in AI usage and analysis of barriers to women's engagement with technology and extract insights.

**Proposition 1:** Women's self-efficacy toward technology must be improved using the AEM model to cross the efficacy-women's-learning-barriers threshold.   …… (P1)

**Proposition 2:** (P2) Women's education must accommodate women's ways of learning and knowing. …… (P2)

**Proposition 3:** P1 & P2 must be simultaneously implemented in any method that seeks to cross the efficacy-women's-learning-barriers threshold  …… (P3)

Proposition 3b: P3 can also be restated as the "Advancement" success zone being positioned at the integrated implementation of P1 & P2 per the AEM model. We have reserved more elaborate discussions on the nature of the barriers and the crossover thresholds for future research, which is purposed using empirical data and experimental studies to measure real impact of implementation self-efficacy improvement measures and women sensitive educational frameworks which cater to women's ways of knowing and gender optimized pedagogy.

**Discussion & Conclusion**

The progress seen to date indicates that the situation can be remedied by taking appropriate steps. This paper presents fresh opportunities for further research by presenting the AEM and AWT model and framework respectively. The research is weak in primary data analysis and we expect to address that issue in the future development of this stream of research. We intend to extent this exploratory analysis using experimental studies, case studies and expert surveys to further test and validate our proposed models. Our propositions while fairly robust from an inductive logic perspective, need to be adapted for hypothesis testing which is expected to provide additional insights.

The research is unique and valuable as it seeks to provide insights into emerging dynamics of education in technology and professional engagement for women. Practitioners can gain actionable insights from the present stream of research to ensure that women do not lose jobs or economic and social opportunities due to incorrect ways of gender engagement. It is possible that more women will lose jobs in technology than men in the forthcoming AI wave driven job loss and replacement of human labor by robotic labor. The present research thus provides some thought provoking insights and basis for further discussion and development.